\documentclass[10pt,aps,prl,twocolumn,superscriptaddress]{revtex4-1}
\usepackage{natbib}
\usepackage[dvips]{graphicx}
\usepackage{bm}
\usepackage{amsmath, amssymb}
\usepackage{upgreek}
\usepackage{url}

\newcommand{\BFA}{BaFe$_{2}$As$_{2}$}
\newcommand{\BFCA}{Ba(Fe$_{1-x}$Co$_{x}$)$_{2}$As$_{2}$}
\newcommand{\KFA}{KFe$_{2}$As$_{2}$}
\newcommand{\BFAP}{BaFe$_{2}$(As$_{1-x}$P$_{x}$)$_{2}$}
\newcommand{\BKFA}{Ba$_{1-x}$K$_{x}$Fe$_{2}$As$_{2}$}

\newcommand{\BFP}{BaFe$_{2}$P$_{2}$}

\newcommand{\Ts}{$T_{\mathrm{s}}$}
\newcommand{\Tc}{$T_{\mathrm{c}}$}
\newcommand{\cm}{cm$^{-1}$}

\newcommand{\ND}{$N_{\mathrm{D}}$}

\newcommand{\LSCO}{La$_{2-x}$Sr$_{x}$CuO$_{4}$}

\begin{document}

\preprint{ver.\ 5.2}

\title{Normal-state charge dynamics in doped \BFA{}: \\ Roles of doping and necessary ingredients for superconductivity}



\author{M.~Nakajima}
\email[]{m-nakajima@aist.go.jp}
\affiliation{Department of Physics, University of Tokyo, Tokyo 113-0033, Japan}
\affiliation{National Institute of Advanced Industrial Science and Technology, Tsukuba 305-8568, Japan}
\affiliation{JST, Transformative Research-Project on Iron Pnictides (TRIP), Tokyo 102-0075, Japan}
\author{S.~Ishida}
\affiliation{Department of Physics, University of Tokyo, Tokyo 113-0033, Japan}
\affiliation{National Institute of Advanced Industrial Science and Technology, Tsukuba 305-8568, Japan}
\affiliation{JST, Transformative Research-Project on Iron Pnictides (TRIP), Tokyo 102-0075, Japan}
\author{T.~Tanaka}
\affiliation{Department of Physics, University of Tokyo, Tokyo 113-0033, Japan}
\affiliation{JST, Transformative Research-Project on Iron Pnictides (TRIP), Tokyo 102-0075, Japan}
\author{K.~Kihou}
\affiliation{National Institute of Advanced Industrial Science and Technology, Tsukuba 305-8568, Japan}
\affiliation{JST, Transformative Research-Project on Iron Pnictides (TRIP), Tokyo 102-0075, Japan}
\author{Y.~Tomioka}
\affiliation{National Institute of Advanced Industrial Science and Technology, Tsukuba 305-8568, Japan}
\affiliation{JST, Transformative Research-Project on Iron Pnictides (TRIP), Tokyo 102-0075, Japan}
\author{T.~Saito}
\affiliation{Department of Physics, Chiba University, Chiba 263-8522, Japan}
\author{C.~H.~Lee}
\affiliation{National Institute of Advanced Industrial Science and Technology, Tsukuba 305-8568, Japan}
\affiliation{JST, Transformative Research-Project on Iron Pnictides (TRIP), Tokyo 102-0075, Japan}
\author{H.~Fukazawa}
\affiliation{JST, Transformative Research-Project on Iron Pnictides (TRIP), Tokyo 102-0075, Japan}
\affiliation{Department of Physics, Chiba University, Chiba 263-8522, Japan}
\author{Y.~Kohori}
\affiliation{JST, Transformative Research-Project on Iron Pnictides (TRIP), Tokyo 102-0075, Japan}
\affiliation{Department of Physics, Chiba University, Chiba 263-8522, Japan}
\author{T.~Kakeshita}
\affiliation{Department of Physics, University of Tokyo, Tokyo 113-0033, Japan}
\affiliation{JST, Transformative Research-Project on Iron Pnictides (TRIP), Tokyo 102-0075, Japan}
\author{A.~Iyo}
\affiliation{National Institute of Advanced Industrial Science and Technology, Tsukuba 305-8568, Japan}
\affiliation{JST, Transformative Research-Project on Iron Pnictides (TRIP), Tokyo 102-0075, Japan}
\author{T.~Ito}
\affiliation{National Institute of Advanced Industrial Science and Technology, Tsukuba 305-8568, Japan}
\affiliation{JST, Transformative Research-Project on Iron Pnictides (TRIP), Tokyo 102-0075, Japan}
\author{H.~Eisaki}
\affiliation{National Institute of Advanced Industrial Science and Technology, Tsukuba 305-8568, Japan}
\affiliation{JST, Transformative Research-Project on Iron Pnictides (TRIP), Tokyo 102-0075, Japan}
\author{S.~Uchida}
\affiliation{Department of Physics, University of Tokyo, Tokyo 113-0033, Japan}
\affiliation{JST, Transformative Research-Project on Iron Pnictides (TRIP), Tokyo 102-0075, Japan}
%\thanks{}


\date{\today}

\begin{abstract}

We carried out a comparative study of the in-plane resistivity and optical spectrum of doped \BFA{} and investigated the doping evolution of the charge dynamics. For \BFA{}, charge dynamics is incoherent at high temperatures. Electron (Co) and isovalent (P) doping into \BFA{} increase coherence of the system and transform the incoherent charge dynamics into highly coherent one. On the other hand, charge dynamics remains incoherent for hole (K) doping. It is found in common with any type of doping that superconductivity with high transition temperature emerges when the normal-state charge dynamics maintains incoherence and when the resistivity associated with the coherent channel exhibits dominant temperature-linear dependence. %The results suggest that doping also controls electronic correlations via chemical pressure and/or change in electron filling, which are responsible for superconductivity in iron arsenides.

\end{abstract}

%\keywords{}

\maketitle

High-transition-temperature (high-\Tc{}) superconductivity in iron arsenides is induced by various doping (chemical substitution) processes into metallic parent compounds showing an antiferromagnetic-orthorhombic (AFO) order. For a representative parent iron arsenide \BFA{}, the superconducting (SC) phase is reached by doping into all the three different lattice sites \cite{Rotter2008,Nandi2010,Kasahara2010}, and even by isovalent atomic substitution, \textit{e.g.}, P for As \cite{Kasahara2010} and Ru for Fe \cite{Thaler2010}. Notably, both electron and hole doping are possible to generate superconductivity, in stark contrast to the case of high-\Tc{} cuprates, in which only one type of carrier doping is possible to induce superconductivity for one parent compound, and the Ni or Zn substitution for the Cu sites never leads to superconductivity.

The mechanism of various doping processes leading to the emergence of superconductivity is an issue of debate. Charge carriers appear to be supplied as revealed by the study of angle-resolved photoemission spectroscopy (ARPES), which demonstrates that the electron Fermi surface (FS) expands (the hole FS shrinks) for electron doping (\textit{e.g.}, Co substitution for Fe) \cite{Liu2010,Ideta2012} and vice versa for hole doping (\textit{e.g.}, K substitution for Ba) \cite{Malaeb2012}. However, it has not been seriously studied how such contrasting doping evolution of the FS is reflected in charge transport. More fundamentally, it is non-trivial how a change in electron/hole density in already metallic parent compounds alters the electronic environment to favor superconductivity, or whether the change in carrier density is really an important ingredient for superconductivity. In addition to the change in carrier density, a chemical-pressure effect also has a significant influence on the electronic structure \cite{Miyake2010a,Usui2011,Nakajima2012b}. To get deeper understanding of the physics of doping in iron-based superconductors, it is of fundamental importance to extract universal natures such as necessary ingredients for superconductivity among different doping processes. Since doping into \BFA{} covers the entire doping range from the AFO to the SC phase and further to the metallic non-SC end compounds, it is a suitable system to investigate the physical implication of doping.

The optical spectroscopy is a bulk-sensitive and energy-resolved probe and turned out to be a powerful tool to study charge dynamics and its evolution with doping in a multi-carrier system, particularly iron pnictides. The optical conductivity spectrum of iron arsenides is characterized by a small zero-energy peak, corresponding to a coherent carrier dynamics, and a long tail extended to a high-energy region, indicative of the presence of an incoherent charge dynamics with large spectral weight \cite{Hu2008,Nakajima2010,Hancock2010,Tu2010,Barisic2010}. Such a spectral feature is also seen in the spectrum of the high-\Tc{} cuprates and characterizes the normal-state charge dynamics in high-\Tc{} superconductors. Combination of resistivity and optical conductivity makes it possible to separate the contribution of multiple carriers or to separate a coherent component from incoherent one in charge dynamics by disentangling the carrier scattering rate from the carrier density. Indeed, the decomposition of the spectrum for Co-doped \BFA{} well explains the temperature and doping dependence of the charge dynamics \cite{Nakajima2010}.

\begin{figure*}
\includegraphics[width=130mm,clip]{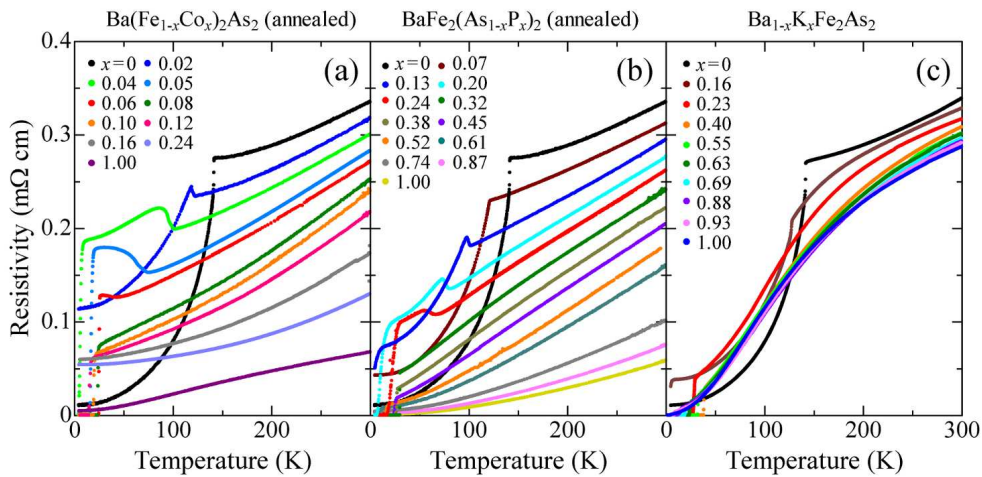}%
\caption{Doping evolution of temperature dependence of the in-plane resistivity for (a) Co-Ba122, (b) P-Ba122, and (c) K-Ba122.}
\end{figure*}

In this paper, we report how the magnitude and the temperature ($T$) dependence of the in-plane resistivity $\rho(T)$ evolve with doping (chemical substitution) into the three different sites in the parent compound \BFA{}: \BFCA{}, \BFAP{}, and \BKFA{} (abbreviated as Co-Ba122, P-Ba122, and K-Ba122, respectively). In the present study, we focus on the effect of doping in the high-temperature paramagnetic-tetragonal (PT) phase. The results of the doping evolution of resistivity and optical conductivity spectrum indicate that a dominant effect is to control coherence/incoherence in charge dynamics in different ways depending on the dopant sites/species. Superconductivity is found to emerge when the normal-state charge dynamics maintains incoherence. It is also found from the analysis of the $T$ dependence of resistivity that, in the region of superconductivity showing fairly high \Tc{} values, the normal-state resistivity is dominated by the $T$-linear inelastic scattering term.

%\section{Experimental}

Single crystals of Co-, P-, and K-doped \BFA{} were grown by the self-flux method \cite{Nakajima2010,Nakajima2012a,Kihou2010}, which were post-annealed to improve the sample quality. Resistivity measurements were performed on the \textit{ab} plane using the four-terminal method. Optical reflectivity was measured at $T=300$ K with the incident light almost normal to the \textit{ab} plane in the frequency range of 50--40000 \cm{} using a Fourier transform infrared spectrometer (Bruker IFS113v) and a grating monochromator (JASCO CT-25C). The optical conductivity was derived from the Kramers-Kronig transformation of the reflectivity spectrum. The Hagen-Rubens or Drude-Lorentz formula was used for the low-energy extrapolation in order to smoothly connect to the spectrum in the measured region and to fit the measured resistivity value at $\omega=0$.

%\section{Results}

%\section{Doping evolution of in-plane resistivity}

Figure 1 shows the doping evolution of $\rho(T)$ for the three systems, Co-, P-, and K-Ba122. The undoped parent compound \BFA{}, the quality of which is remarkably improved by annealing, shows very low residual resistivity ($<$ 10 $\upmu\Omega$\,cm) at low temperatures in the AFO phase, and the AFO-PT transition temperature \Ts{} is 143 K, the highest among so far reported \cite{Nakajima2011,Ishida2011}. Nevertheless, the resistivity at 300 K is fairly high, $\sim$ 340 $\upmu\Omega$\,cm and is comparable with that for unannealed crystals. The $T$ dependence is weak in the PT phase above \Ts{}, suggesting that charge carriers are highly incoherent. The resistivity in the PT phase rapidly decreases with Co and P doping, in sharp contrast with that in the AFO phase, where the resistivity increases with doping, indicating different roles of doping between the two phases. By contrast, the change in magnitude is moderate in the case of K doping.

%\subsection{\BFCA{}}
With Co doping, the temperature dependence changes from $T$ linear to $T^2$ in the overdoped non-SC region. Note that a fairly large residual resistivity remains up to $x=0.24$, indicating that a doped Co atom works as a relatively stronger scattering center. %The end compound \BCA{} is a good metal with low resistivity, $\rho$(300 K) $\sim$ 70 $\upmu\Omega$\,cm and $\rho$(5 K) $\sim$ 5 $\upmu\Omega$\,cm, which is in quite contrast to the bad-metallic behavior of \BFA{}. The evolution of $\rho$ with Co doping could be understood if a doped Co atom adds one electron carrier to the system. %In fact, angle-resolved photoemission spectroscopy (ARPES) measurements have revealed the doping evolution of Fermi surface (FS) in accordance with a rigid-band model \cite{Liu2010}, and the Hall coefficient indicates that the electron density increases with $x$ \cite{Albenque2009}.
%\subsection{\BFAP{}}
The substitution of P for As corresponds to chemically isovalent doping. The number of electrons and holes is expected to be balanced over the entire range of doping. %Indeed, the overall feature of both electron and hole FS pockets does not change markedly over the entire P concentration (the FS volume weakly increases, and one of the hole FS pockets is more warped along the $c$ axis \cite{Ye2011}). 
Controversially, as in the case of Co-Ba122, the magnitude of the resistivity in the PT phase monotonically decreases with P doping, and the $T$ dependence gradually changes from $T$-linear to $T^2$, in agreement with the result by Kasahara \textit{et al.}\ \cite{Kasahara2010}. A difference is in that a doped P atom is a weaker scattering center as evidenced by a smaller residual-resistivity component. %a tendency for resistivity saturation is visible for the compounds with the P content $x \le 0.45$ at high temperatures. Like \BCA{}, the end material \BFP{} is a good metal [$\rho$(300 K) $\sim$ 55 $\upmu\Omega$\,cm and $\rho$(5 K) $<$ 1 $\upmu\Omega$\,cm].
%\subsection{\BKFA{}}
The doping evolution of $\rho(T)$ in K-Ba122 is in quite contrast to the Co- and P-doping cases. Chemically, K doping corresponds to hole doping, but both magnitude and $T$ dependence of the resistivity only weakly change with doping. %and indeed the ARPES experiment observed an expansion of hole FS pockets (a shrinkage of electron FS pockets) \cite{Malaeb2012}. 
At low temperatures and at any doping level, the resistivity rapidly decreases with decreasing temperature, showing $T^2$ dependence, typical of a Fermi liquid, whereas it shows a clear tendency for saturation in the high-temperature PT phase. As far as the high-$T$ region is concerned, even the end compound \KFA{} with \Tc{} $\sim$ 3.5 K is not a good metal. A finite residual resistivity appears for the compounds showing the AFO order ($x \le 0.23$), but, for $x \ge 0.40$, it is extremely small probably because a doped K atom is located farthest away from the Fe plane. %showing high resistivity and weak $T$ dependence. Remarkably, the magnitude and the $T$ dependence of resistivity in the PT phase no more change appreciably in the overdoped regime ($x > 0.40$), whereas \Tc{} decreases with increasing K content. This is in apparent contradiction to the ARPES result, which shows the systematic development of the hole FS with doping \cite{Malaeb2012}.

%\subsection{Doping evolution}

Figure 2(a) displays the optical conductivity spectrum of \BFA{}. The spectrum shows a small peak at $\omega=0$ and a long tail merging into a broad peak at higher-energy region. Using the Drude-Lorentz model, we decompose the spectrum as shown in Fig.\ 2(a) \cite{Nakajima2010,Wu2010}. The black and orange lines indicate the experimental and fitting results, respectively. The low-energy conductivity is dominated by two Drude components with distinct characters; one is narrow with small spectral weight (coherent Drude component $\sigma_{\mathrm{D}}$, in blue) and the other is broad with much larger spectral weight (incoherent component $\sigma_{\mathrm{in}}$, in gray),
\begin{equation}
\sigma(T, \omega) = \sigma_{\mathrm{D}}(T, \omega) + \sigma_{\mathrm{in}}(\omega).
\end{equation}
A higher-energy component approximated by a Lorentzian function (in green) corresponds to an interband excitation.

\begin{figure}
\includegraphics[width=83mm,clip]{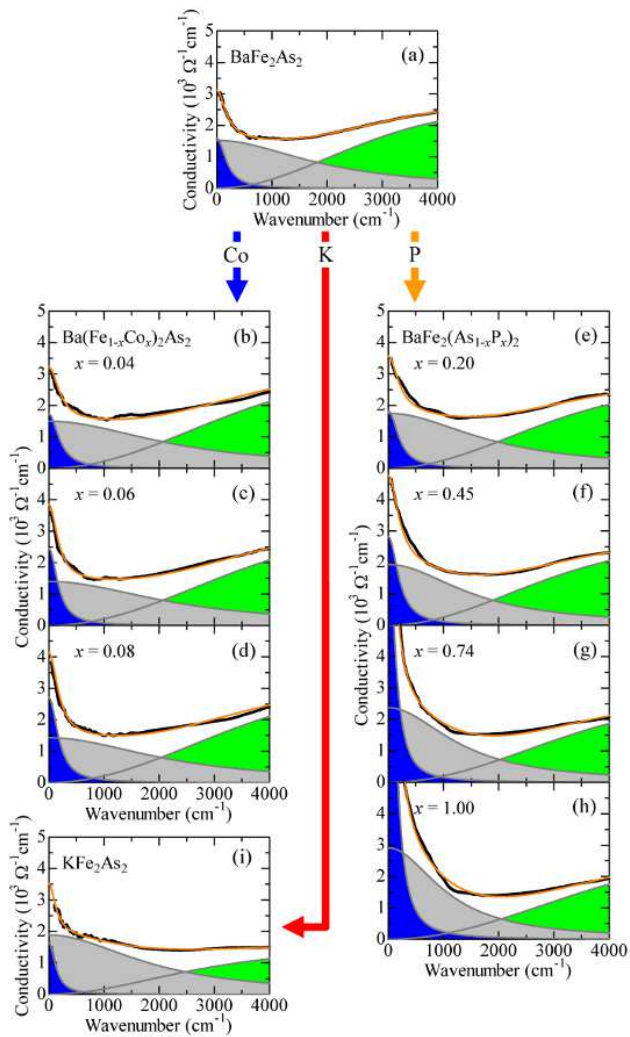}%
\caption{Decomposition of the optical conductivity spectrum at $T=300$ K for (a) \BFA{}, (b--d) Co-Ba122, (e--h) P-Ba122, and (i) \KFA{}.}
\end{figure}

The evolution of each component with $x$ for Co-Ba122 is shown in Figs.\ 2(b)--2(d) \cite{Nakajima2010}. With Co doping, the weight of the narrow Drude component (coherent Drude weight) increases, whereas that of the broad Drude component shows no appreciable change. %For the end compound \BCA{} [Fig.\ 3(b)], the narrow Drude weight increases by a factor of 6 compared with that of \BFA{}, and even the broad Drude term is significantly narrowed in width, indicating that most of carriers gain coherence.
Figures 2(e)--2(h) show the evolution of the spectrum with P doping. The substitution of isovalent P for As supplies neither extra electrons nor holes. However, the coherent Drude weight remarkably increases with P doping without changing the total low-energy spectral weight below 6000 \cm{}---the spectral weight is transferred to the coherent component from higher-energy region \cite{Nakajima2012b}. As a result, the narrow Drude component dominates in the low-energy spectrum of \BFP{} [Fig.\ 2(h)].

The decomposition of the spectrum of \KFA{} [Fig.\ 2(i)] is nearly the same as that of \BFA{}, in which the fraction of the coherent Drude weight remains small, or even smaller, and hence the low-energy conductivity spectrum is also dominated by the incoherent component. In view of the spectrum presented in Ref.\ \onlinecite{Li2008} for the K content $x=0.4$, the highly incoherent spectrum seems to persist over the entire doping range of K-Ba122 without showing any appreciable spectral change.

%The much reduced scattering rates result from Fermi-surface reconstruction associated with the AFO order in \BFA{} and a radical suppression of the inelastic scattering process in \KFA{}.

%\subsection{Electron-hole asymmetry}

Here, we discuss the results of resistivity and optical conductivity for the three doped \BFA{} systems in relation with the available results of the Hall coefficient and the FS observed by ARPES. In the high-temperature PT phase of \BFA{}, there are two electron FS pockets near the Brillouin zone boundary and three hole FS pockets around the zone center. The Co doping makes the hole FS shrink and the electron FS expand \cite{Liu2010,Ideta2012}. The Hall coefficient, indeed, decreases in magnitude with negative sign as doping proceeds \cite{Albenque2009}. Hence, the increase in the coherent Drude weight with Co doping is associated predominantly with the increase in electron carriers supplied by the dopant Co atoms, not localized on the Co atom \cite{Wadati2010}.

On the other hand, the hole FS expands (the electron FS shrinks) with K doping, and eventually the electron FS disappears in the highly doped region \cite{Malaeb2012}. However, the Hall coefficient in the PT phase does not appreciably change over a wide range of hole doping \cite{Shen2011,Ohgushi2012}. Consistent with the result of the Hall effect, the resistivity does not decrease appreciably with K doping, and the coherent Drude weight remains small over the entire doping range. These are in contrast to the electron-doped Co-Ba122 system. To reconcile with the doping evolution of the FS observed by ARPES, it is necessary to assume that quasiparticle states on a substantial part of the hole FS pockets remain incoherent in the hole-doped system.

The result for the isovalent P doping gives an additional evidence for the electron-hole asymmetry. With P doping, the numbers of the electron and hole FS pockets do not change and thus the FS topology does not significantly change over the entire range of P content, although one of the hole FS pockets is more warped along the $c$-axis direction as $x$ increases \cite{Ye2011}. Nevertheless, the P doping decreases the resistivity and transforms the bad metallic \BFA{} into a good metal as in the case of Co doping. The P doping increases the coherent Drude weight without changing the total low-energy spectral weight \cite{Nakajima2012b}. In view of the negative Hall coefficient, the magnitude of which decreases with increasing P content \cite{Kasahara2010}, it would follow that, as doping proceeds, the coherent region in the momentum space gradually expands within the electron FS pockets, while the states on the hole FS pockets remain incoherent. This reconciles with the conclusion drawn from the contrasting evolution of coherence/incoherence for electron- and hole-doped \BFA{}. The result of P-Ba122 is reminiscent of the hole-doped high-\Tc{} cuprates, in which hole doping makes the coherent states extend from the nodal to the antinodal region on the FS without changing the FS topology.

%\subsection{Degree of coherence and electronic correlations}

\begin{figure}
\includegraphics[width=65mm,clip]{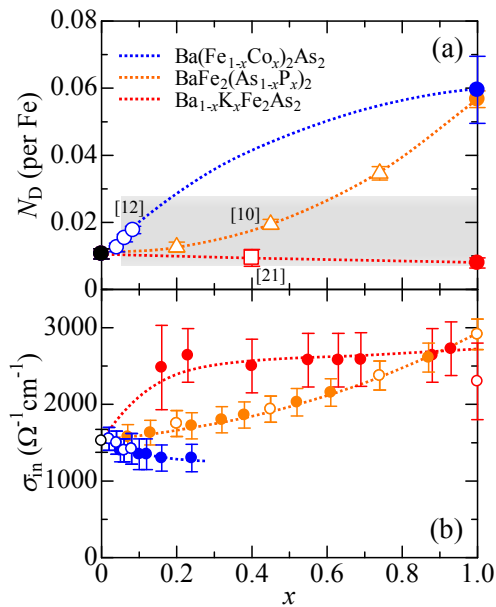}%
\caption{(a) Spectral weight of the narrow Drude component \ND{} as a function of element substitution content $x$. Co-Ba122, P-Ba122, and K-Ba122 are indicated in blue, orange, and red curves, respectively. Open symbols show the data from Refs.\ \onlinecite{Nakajima2010,Nakajima2012b,Li2008}. (b) $\omega=0$ value of the broad Drude component as a function of $x$. Open and closed symbols show the values obtained from the decomposition of the optical conductivity spectrum (Fig.\ 2) and from the analysis of the temperature dependence of resistivity, respectively.}
\end{figure}

In Fig.\ 3(a), we plot the weight of the narrow Drude component \ND{} as a function of the dopant concentration $x$ for the three systems. Open symbols indicate the data compiled from Refs.\ \onlinecite{Nakajima2010,Nakajima2012b} and \onlinecite{Li2008}. Strikingly different evolutions of \ND{} are seen, despite the similar phase diagram for each system. For Co-Ba122, \ND{} increases \textit{sublinearly} with $x$, a steep initial increase but eventually slowing down. \ND{} of P-Ba122 shows a \textit{superlinear} increase with $x$, a gradual increase followed by a rapid increase for $x>0.6$. The much faster increase in \ND{} with Co doping indicates that electron carriers are involved in the coherent charge dynamics. By contrast, \ND{} for K-Ba122 weakly decreases with increasing $x$. Compared with the results of resistivity shown in Fig.\ 1, it is clear that doping dependence of \ND{} traces the doping dependence of resistivity in the high-$T$ region in each system. Note that, in the case of K doping, the resistivity only slightly decreases with $x$, but the hole density steadily increases, as evidenced by the increase in the volume of the hole FS observed by ARPES. The measurement of quantum oscillations for \KFA{} reported an enhancement of the carrier effective mass $m^*$ by a factor of 4--5 as compared with that in \BFA{} \cite{Terashima2010,Terashima2011}, indicating that the decrease in \ND{} is likely due to the increase in $m^*$. Probably, the coherent carrier density $n_{\mathrm{D}}$ would increase with a rate of increase slower than that of $m^*$ (\ND{} $= \frac{m_0}{m^*} n_{\mathrm{D}}$; $m_0$ and $n_{\mathrm{D}}$ being the free electron mass and the coherent carrier density, respectively).

Contrary to the K doping, a decrease in $m^*$ with $x$ (for $x>0.3$) is reported for P doping by quantum oscillations \cite{Shishido2010}. $m^*$ decreases by a factor of 2--3 from $x=0.3$ to 1. In this respect, the increase in \ND{} for P-Ba122 may, in most part, be due to a decrease in $m^*$. %For the Co-doped system, there is no data at the moment which indicates a change in $m^*$, but it is likely that, in the PT phase, in addition to the increase in coherent electron density, a decrease in $m^*$ contributes to the increase in \ND{} to some extent.

%The broad Drude component is also affected by doping. Figure 3(b) shows the doping dependence of the $\omega=0$ conductivity contributed by the broad Drude component ($\sigma_{\mathrm{in}}$). With Co doping, $\sigma_{\mathrm{in}}$ slightly decreases. For P-Ba122, $\sigma_{\mathrm{in}}$ monotonically increases with doping. This stems from the narrowing of the broad Drude component, pointing toward the coherent charge dynamics. With K doping, $\sigma_{\mathrm{in}}$ initially shows a rapid increase and keeps a nearly constant large value for a wide range of doping, which leads to highly incoherent charge dynamics. Such a large value of $\sigma_{\mathrm{in}}$ is responsible for the high resistivity and the tendency of saturation at high temperatures.

%\section{Necessary ingredients for high-\Tc{} superconductivity}

%\subsection{Bad metallic behavior in the normal state}

A bad-metallic behavior in the iron arsenides manifests in the high-$T$ region, where the magnitude of resistivity is fairly high, and the resistivity shows a tendency for saturation. It signals a dominant contribution of the broad (incoherent) Drude component at high temperatures in Eq.\ (1). $\sigma_{\mathrm{in}}$ is only weakly dependent on temperature, which is found to well reproduce the low-energy $\sigma(T, \omega)$ \cite{Nakajima2010}. The resistivity saturation is understood in terms of conductivity in two channels at $\omega=0$,
\begin{equation}
\rho^{-1}(T) = \sigma(T) = \sigma_{\mathrm{D}}(T) + \sigma_{\mathrm{in}},
\end{equation}
where $\sigma_{\mathrm{in}}$ is the $\omega=0$ value of the incoherent conductivity term. For iron arsenides, $\sigma_{\mathrm{in}}$ is comparable with $\sigma_{\mathrm{D}}$. The resistivity from the coherent component is expressed by a sum of the contributions from $T$-independent elastic and $T$-dependent inelastic scattering ($\rho_0$ and $\rho_{\varepsilon}$, respectively), $\sigma_{\mathrm{D}}^{-1}(T) = \rho_0 + \rho_{\varepsilon}(T)$. With increasing temperature, $\sigma_{\mathrm{D}}$ decreases, and consequently the contribution of $\sigma_{\mathrm{in}}$ becomes dominant.

Eq.\ (2) happens to be similar to a simple parallel resistor formula empirically describing the resistivity saturation, in which $\sigma_{\mathrm{in}}$ is replaced by $\rho_{\mathrm{sat}}^{-1}$, $\rho_{\mathrm{sat}}$ being the maximum resistivity corresponding to the carrier mean free path as short as the lattice spacing \cite{Wiesmann1977}, $\rho^{-1}(T) = \sigma_{\mathrm{D}}(T) + \rho_{\mathrm{sat}}^{-1}$. The present $\sigma_{\mathrm{in}}$ is not such a universal quantity but is dependent on material and doping level as shown in Fig.\ 3(b). This is distinct not only from the case of conventional resistivity saturating metals but also from the non-saturating resistivity of the known bad metals, many of which are in close proximity to a Mott insulating state such as high-\Tc{} cuprates \cite{Gunnarsson2003,Hussey2004}. The non-saturating resistivity is thought to arise from a reduction of low-frequency conductivity with elevating temperature, the spectral weight of which is transferred to higher frequencies involving transitions to upper Hubbard band \cite{Hussey2004,Deng2012}. It is likely that the presence of multiple carriers with distinct characters is responsible for the apparent saturation behavior of the iron arsenides.

%In Fig.\ 3(a), we plot the coherent Drude weight \ND{} as a function of the dopant concentration $x$ for the three systems. Open symbols indicate the data compiled from Refs.\ \onlinecite{Nakajima2010,Nakajima2012b} and \onlinecite{Li2008}. Strikingly different evolutions of \ND{} are seen, despite the similar phase diagram for each system.
The result shown in Fig.\ 3(a) demonstrates that doping controls \ND{} in quite different manners depending on dopant sites and/or species. It is surprising that either of such different doping processes leads to the emergence of superconductivity. %(As shown in a separate paper, the reduction rate of the AFO ordering temperature, $-\frac{\mathrm{d}T_{\mathrm{s}}}{\mathrm{d}x}$, is largely different among these three doping processes, largest for Co doping and smallest for K doping, just in the order of decreasing strength of disorder introduced by doping.)
%The SC phase appears irrespective of the rate of the suppression of the AFO order \cite{Ishida2012b}. 
The critical doping level above which superconductivity appears more or less depends on the suppression rate of the AFO order, but it is open what conditions should be fulfilled for superconductivity to appear. %The width of the SC dome, the $x$ range in which superconductivity shows up, expands also in the order of Co, P, and K doping. 
On the other hand, the result indicates that the SC phase persists as long as \ND{} is sufficiently small [\ND{} $<$ 0.02--0.03, shaded region in Fig.\ 3(a)] and that the system cannot sustain superconductivity when \ND{} gets larger or charge dynamics becomes more coherent. \ND{} rapidly increases with Co doping, and soon reaches $\sim 0.02$ at $x \sim 0.10$. In the case of P doping, \ND{} increases slowly and exceeds 0.03 at around $x=0.6$. This seems to corroborate the fact that the SC phase terminates at $x \sim 0.15$ and $\sim 0.7$ for Co and P doping, respectively. Different from these two systems, \ND{} of K-Ba122 maintains small values up to $x=1$, consistent with the persistent superconductivity to \KFA{}. %In agreement with this result, the correlation diagram shown in Fig.\ 6 indicates that superconductivity with fairly high \Tc{} is realized when the electronic correlation is not too weak nor too strong.
The incoherent normal-state charge dynamics is therefore a prerequisite for superconductivity.

%\subsection{$T$-linear resistivity in high-\Tc{} regime}

\begin{figure}
\includegraphics[width=80mm,clip]{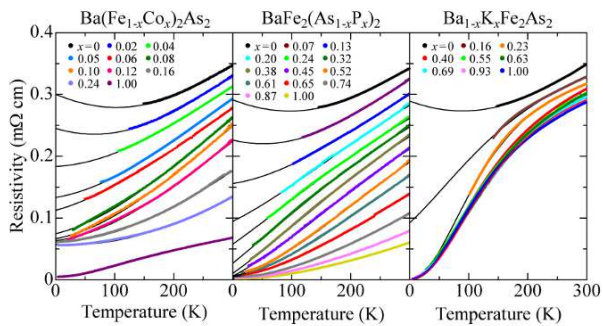}%
\caption{Fitting of the temperature dependence of resistivity for Co-, P-, and K-Ba122 in the PT phase using a formula $1/\rho(T) = 1/(\rho_0 + \alpha_1 T + \alpha_2 T^2) + \sigma_{\mathrm{in}}$.}
\end{figure}

Finally, we analyze the $T$ dependence of the resistivity in the PT phase predominantly arising from the coherent component and search for correlation with superconductivity. Assuming the two parallel (inelastic) scattering channels, we adopt a dual-component analysis for the $T$-dependent part of resistivity, $\rho_{\varepsilon}(T) = \alpha_1 T + \alpha_2 T^2$. The $T^2$ term is typical of a conventional Fermi liquid, and the $T$-linear one usually results from electron-boson interactions such as antiferromagnetic spin fluctuations. This dual-component analysis was applied to fit the resistivity of cuprates \cite{Cooper2009} and iron pnictides \cite{Leyraud2009}. However, this formula does not reproduce a trend for saturation observed for most of iron arsenides at high temperatures. Therefore, we have to incorporate the incoherent channel $\sigma_{\mathrm{in}}$ explicitly and use a formula
\begin{equation}
1/\rho(T) = 1/(\rho_0 + \alpha_1 T + \alpha_2 T^2) + \sigma_{\mathrm{in}},
\end{equation}
where $\rho_0$ represents the elastic scattering component adding to the inelastic scattering processes. $\sigma_{\mathrm{in}}$ can be estimated from the $\omega \rightarrow 0$ value of the broad (incoherent) Drude component in the present optical conductivity spectrum \cite{sigma_in}. Figure 4 shows the fitting results of the resistivity curves in the PT phase for the present three systems. Formula (3) with $\sigma_{\mathrm{in}}$ well reproduces the resistivity curves over wide temperature and doping range.

\begin{figure*}
\includegraphics[width=150mm,clip]{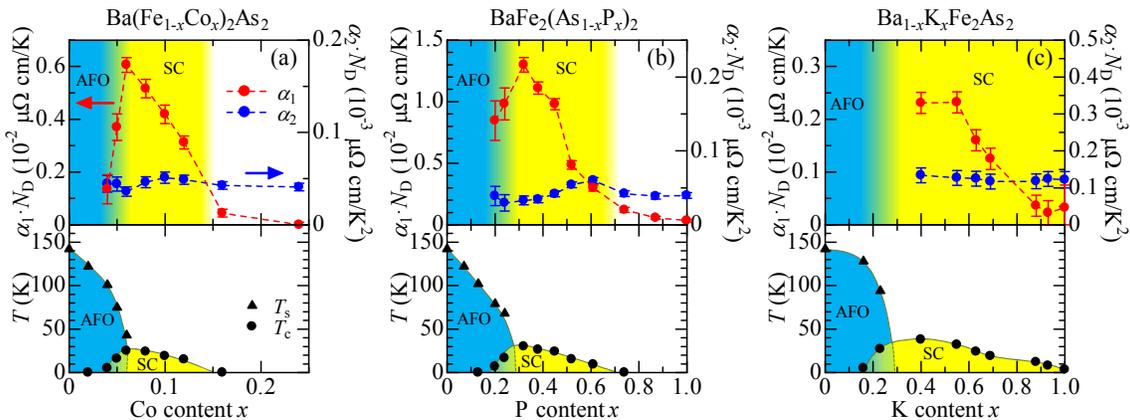}%
\caption{Doping evolution of the coefficients of the $T$-linear and $T^2$ components ($\alpha_1 \cdot$\ND{} and $\alpha_2 \cdot$\ND{}, respectively) for (a) Co-Ba122, (b) P-Ba122, and (c) K-Ba122 determined from the fitting shown in Fig.\ 4. The $T=0$ electronic phases are color-coded: AFO in blue, SC in yellow, AFO-SC coexisting region in gradation, and non-SC metallic in white. Lower panels show the electronic phase diagrams for each system.}
\end{figure*}

Figures 5(a)--5(c) show the doping dependences of the coefficients of the $T$-linear and square terms multiplied by \ND{}, $\alpha_1 \cdot$\ND{} and $\alpha_2 \cdot$\ND{}, respectively, for the three doping systems obtained using the best-fit parameters, $\alpha_1$ and $\alpha_2$, to reproduce the resistivity curves (Fig.\ 4). For the compositions we did not measure the optical spectrum, \ND{} is estimated by assuming smooth variation with $x$ shown as the dashed curves in Fig.\ 3(a). In the framework of the present two-fluid and dual-channel analysis, $\alpha_1 \cdot$\ND{} and $\alpha_2 \cdot$\ND{} are more directly related with the scattering rate of the coherent carriers $1/\tau_{\mathrm{D}}$ ($\sigma_{\mathrm{D}} = N_{\mathrm{D}} e^2 \tau_{\mathrm{D}} / m_0$). In common with the three systems, $\alpha_1 \cdot$\ND{} is largest near the AFO-SC phase boundary where \Tc{} is highest, whereas it becomes vanishingly small in the overdoped regime where superconductivity disappears in Co- and P-Ba122. Superconductivity in K-Ba122 persists up to $x=1$, but $\alpha_1 \cdot$\ND{} remarkably decreases on going from $x=0.69$ to 0.88. This probably corresponds to the disappearance of the electron FS pockets \cite{Malaeb2012} and/or the change in the SC gap function on FS from nodeless \cite{Ding2008} to nodal one \cite{Fukazawa2009,Okazaki2012}. Unlike $\alpha_1 \cdot$\ND{}, $\alpha_2 \cdot$\ND{} remains finite over a wide range of doping regardless of the \Tc{} value and does not show any remarkable feature at the phase boundary. As displayed in Fig.\ 3(a), the weight of the coherent Drude component \ND{} monotonically varies with $x$. Therefore, the non-monotonic change in $\alpha_1 \cdot$\ND{} certainly arises from that of $\alpha_1$, the $T$-linear channel of the carrier scattering rate.

The predominance of the $T$-linear term implies stronger inelastic scattering of carriers, and the observed correlation between $\alpha_1 \cdot$\ND{} and \Tc{} is highly suggestive of the scattering mechanism giving rise to the $T$-linear resistivity having an intimate connection with the SC-pair formation. Cooper \textit{et al.} attempted to fit the temperature dependence of the in-plane resistivity of highly doped \LSCO{} ($x>0.15$) using a similar formula \cite{Cooper2009}. They found that $\alpha_1$ is largest at $x=0.19$ and decreases with increase or decrease in the doping level, whereas $\alpha_2$ is only weakly dependent on the hole concentration down to $x=0.19$. It should be noted that the $T$-linear resistivity in the hole-doped cuprates is different in nature. $T$-linearity in the cuprates persists up to very high temperatures, $\sim$ 1000 K or even higher without showing resistivity saturation, and originates from highly incoherent quasiparticle states. The quasiparticle scattering rate ($1/\tau$) in the cuprates is strongly momentum dependent and is extremely high at momenta near the antinodes, $\hbar/\tau \sim 2\pi k_{\mathrm{B}}T$, which is near the so-called Planckian dissipation limit \cite{Cooper2009,Zaanen2004}. This is in contrast to the present iron-pnictide case, in which the $T$-linear term arises from the coherent channel in parallel to the incoherent one, and the resistivity exhibits a tendency for saturation due to the nearly $T$-independent incoherent term. The width of the narrow Drude component is $\sim$ 200 \cm{} at $T=300$ K for every system near the doping level where \Tc{} reaches the highest value. This corresponds to $\hbar/\tau \sim k_{\mathrm{B}}T$, by a factor of $\sim$ 6 lower than that in \LSCO{}.

The $T$-linear resistivity is often argued in relation to enhanced scattering from critical fluctuations near a quantum critical point (QCP). For P-Ba122, the presence of a magnetic QCP at $x \sim 0.3$ was suggested \cite{Shishido2010,Hashimoto2012}.
%However, for LSCO and also for P-Ba122, $\alpha_1$ and $\alpha_2$ remain finite over a wide range of $x$, which is not a conventional QCP behavior such as seen in YbRh$_2$Si$_2$ \cite{Custers2003}. Moreover, in the present system, $\alpha_2 \cdot$\ND{} does not show any critical change at $x$ where $\alpha_1 \cdot$\ND{} is peaked. As shown in Fig.\ 6, these features of $\alpha_1 \cdot$\ND{} and $\alpha_2 \cdot$\ND{} are also in common with the electron- and hole-doped \BFA{}. This is also a difference from the cuprates.
The $T$-linear resistivity coefficient $\alpha_1$ is peaked at $x \sim 0.3$, and the effective mass $m^*$ estimated from de Haas oscillations increases from the overdoped region toward $x \sim 0.3$. However, this is specific to the P-doped case. Such mass enhancement is not observed for Co-Ba122, and, to the contrary, $m^*$ increases toward $x=1$ for K-Ba122. Moreover, for every doped system, the SC phase is not confined to a region where $\alpha_1$ (or $\alpha_1 \cdot$\ND{}) shows a peak and persists to $x=1$ in the case of K-Ba122. Therefore, the present results support a scenario that superconductivity in the iron arsenides arises, irrespective of presence or otherwise of a QCP, from strongly $x$-dependent coupling with some bosonic excitations, which is a source of the $T$-linear carrier scattering. The facts that $\alpha_1$ and \Tc{} attain large values near the AFO-SC phase boundary and that they steeply decrease as $x$ goes away from the boundary suggest that the bosonic excitations originate from the fluctuations of the AFO order. As we discussed in the preceding study of the anomalous response of the AFO phase to dopant impurities \cite{Nakajima2012c,Ishida2012}, the AFO phase is a unique spin-charge-orbital complex. Hence, inelastic scattering from short-ranged dynamical spin-charge-orbital correlations is a possible candidate of the $T$-linear resistivity and hence pairing interactions.

%\section{Summary}

To summarize, we investigated the effect of various dopings on the charge dynamics of \BFA{} in the PT phase by the combined measurements of in-plane resistivity and optical conductivity spectrum. \BFA{} is characterized by a bad metallic behavior with a coherent Drude component occupying only a tiny fraction of the low-energy spectral weight in the optical conductivity spectrum. Co and P doping transforms a bad metal to a good metal, while the system remains a bad metal with K doping. %The coherent Drude spectral weight or its fraction in the total low-energy spectral weight remarkably increases with isovalent P doping, which is ascribed to an increase in the bandwidth via chemical pressure associated with an in crease in the As-Fe-As bond angle. The charge dynamics is more coherent and the electronic correlation is weaker with an increase in the bandwidth. The increase in the coherent spectral weight with Co doping is due to the increase in the coherent electron density, and the chemical pressure may also be at work as in the case of P-Ba122. To the contrary, the coherent spectral weight decreases with K doping, which, in part, acts as negative chemical pressure, decreasing the bandwidth and also makes the electron filling decrease toward half filling.
%An inspection of doping dependence of lattice parameters suggests that a parameter controlling coherence/incoherence is the As-Fe-As bond angle, which controls the bandwidth and thereby the strength of electronic correlations.
%Distinct doping evolution of the degree of coherence between hole and electron doping suggests that the band filling also controls correlation.
It is found that, irrespective of the doping type, superconductivity with relatively high \Tc{} is realized in the doping region where the charge dynamics remains incoherent. Thus, the bad metallic normal state or the incoherent charge dynamics is one of the prerequisites for the emergence of high-\Tc{} superconductivity in the \textit{TM} pnictides as in the case of the cuprates. In the high-\Tc{} regime, $T$ dependence of resistivity is dominated by the $T$-linear term. The magnitude of the scattering rate, which gives rise to the $T$-linear term in resistivity, is found to be well correlated with the superconducting \Tc{}. %This finding is strongly suggestive of an intimate connection among incoherent charge dynamics, $T$-linear resistivity, strong electronic correlation, and superconductivity with high \Tc{}.

%\appendix
%\section{}

\begin{acknowledgments}

M.N. and S.I. thank the Japan Society for the Promotion of Science (JSPS) for the financial support. Discussions with T. Misawa, M. Imada, and A. Fujimori were helpful in preparing this manuscript. This work was supported by the Japan-China-Korea A3 Foresight Program and Grant-in-Aid for JSPS Fellows from JSPS, Grant-in-Aid for Scientific Research from JSPS and the Ministry of Education, Culture, Sports, Science, and Technology, Japan, and the Strategic International Collaborative Research Program (SICORP) from Japan Science and Technology Agency.

\end{acknowledgments}


\begin{thebibliography}{00}
\bibitem{sigma_in} For K-Ba122, larger values of $\sigma_{\mathrm{in}}$ are required to fit the resistivity curves than the experimentally extracted values. This might indicate that the decomposition of optical conductivity underestimates the contribution from the incoherent component, pointing toward the more incoherent charge dynamics inherent in K-Ba122 at high temperatures.
\end{thebibliography}
\end{document}